# Entropic topography associated with field-induced quantum criticality in a magnetic insulator DyVO$_4$

Dheeraj Ranaut and K. Mukherjee

School of Basic Sciences, Indian Institute of Technology Mandi, Mandi 175005, Himachal Pradesh, India

**Abstract**

Exploration of low temperature phase transitions associated with quantum critical point is one of the most mystifying fields of research which is under intensive focus in recent times. In this work, through comprehensive experimental evidences, we report the possibility of achieving quantum criticality in the neighborhood of a magnetic field-tuned tricritical point separating paramagnetic, antiferromagnetic and metamagnetic phases in a magnetic insulator, DyVO$_4$. Magnetic susceptibility and heat capacity indicate to the presence of a long-range second order antiferromagnetic transition at $T_N \sim 3.2$ K. Field variation of Magnetic susceptibility and heat capacity, along with differential magnetic susceptibility and DC field dependent AC susceptibility gives evidence of the modification of the antiferromagnetic structure below the tricritical point; implying the presence of a field-induced first order metamagnetic transition which persists down to 1.8 K. Further, the magnetic field dependence of the thermodynamic quantity -d$M$/d$T$, which is related to magnetic Gruneisen parameter, approaches a minimum, followed by a crossover near 5 kOe to a maximum; along with a hyperbolic divergence in temperature response of d$M$/d$T$ in the critical field regime. Temperature response of heat capacity at 5 kOe also shows a deviation from the conventional behavior. Entropic topography phase diagram allows tracking of the variation of the entropy, which indicates towards the emergence of the peak at quantum critical point into a V-shaped region at high temperatures. Our studies yield an inimitable phase diagram describing a tricritical point at which the second-order antiferromagnetic phase line terminates followed by a first order line of metamagnetic transition, as the temperature is lowered, leading to metamagnetic quantum critical end point.

**Introduction**

Over the last few decades, with the development of novel materials, it has become possible to study the unstable phases of matter near the interface between the stable phases. From this point of instability between two stable phases, i.e. quantum critical point (QCP), completely new features in certain materials, arises [1, 2]. Investigations related to QCP associated with quantum phase transitions (QPT) has become one of the fascinating topics in modern condensed matter physics due to rich and enigmatic physics associated with them. Around QCP, quantum mechanical fluctuations play a prominent role in order to achieve continuous phase transitions; analogous to the thermal fluctuations in the classical case [3]. In case of a second order phase transition (SOPT), a QCP can be achieved by the suppression of long-range magnetic ordering to $T = 0$ K ($T$ is the temperature); separating the stable magnetic and non-magnetic phases [1, 3]. The QCP can also be obtained when the SOPT is not continuous and develops into a first order phase transition (FOPT) at a tricritical point (TCP), which is then suppressed to $T = 0$, without violating any symmetry conservation [4]. Although a material can never be cooled down to QCP at absolute zero, drastic effects can be felt long before reaching the QCP [1, 2]. In the region, $T \geq 0$, near the vicinity of QCP, unusual physical properties are expected due to the involvement of low-lying excited states, driven by quantum and thermal fluctuations collectively. In case of metals, quantum criticality is seen to bring qualitative changes in the electronic structure, which is reflected at the Fermi surface. A diverse variety of metallic materials including cuprates, ruthenates, organic metals, and heavy-fermion compounds tuned to QCPs show novel properties like nematic state, strange metal behavior etc, at QCPs [2]. While for insulators, the electron charge is localized, and the study is mainly focused on the orientation of spins on the lattice sites. A rich variety of magnetic phases is possible which are insulating in nature. Among these, the fragile types of antiferromagnetic (AFM) ordering (like staggered, alternating etc) at low temperatures are more susceptible to melting by quantum mechanical fluctuations [2]. As a result, unusual temperature divergences in the magnetic susceptibility and heat capacity are noted [1, 5].

In very few compounds, magnetic ordering persists down to 0 K. In most of the cases, non-thermal parameters like chemical substitution, external pressure, or magnetic field act as driving forces for tuning the system towards QCP [5]. It is generally difficult to perform thermodynamic measurements at high pressures, although such measurements are important for understanding how cooperative phases are stabilized at low temperatures. On the other hand, chemical substitution introduces disorder in the system which makes it difficult to discern its effect from quantum criticality. In this regard, magnetic field appears to be quite useful parameter as it can reversibly and continuously tune the system towards QCP without bothering much about the above-mentioned complexities [6]. In the context of quantum criticality, the thermodynamic properties, especially entropy ($S$) plays an important role. Lijun Zhu *et al*., reported that on approaching a QCP, the $T$ and magnetic field ($H$) variation of $S$ is expected to show non-linear behavior [7]. Particularly, $S$ ($H$) is anticipated by a pronounced peak, expected to be centered at QCP. In magnetic insulators, QCP represents a kind of void in the material phase diagram, and it distorts the curvature of the phase diagram, creating a V-shaped quantum critical region, fanning out from the QCP. This feature is quite visible in the magnetic field-temperature phase diagram; especially in resistivity for metals, and in $S$ for insulators [1, 2]. $CoNb_2O_6$ is a rare example of magnetic insulators showing such kind of behavior in which the QCP is described by the conformal field theory [1, 8]. In addition, a thermodynamic quantity -d$M$/d$T$, which is related to magnetic Gruneisen parameter ($\Gamma_{mag}$), acts as an important probe for the investigation of the QCP [7, 9]. For a system having ordered state at finite $T$, a sign change of the $\Gamma_{mag}$ takes place near QCP which originates from the quantum critical contribution to the thermal fluctuations [9]. Also, for systems having first order line of transition, $\Gamma_{mag}$ show a minimum followed by a crossover near metamagnetic critical field ($H_m$) to a maximum (or vice versa) [9, 10, and 11]. $T$ dependence of this quantity also shows a hyperbolic divergence with decreasing $T$ for $H_m$, which is attributed to the quantum phase transition [12]. Field induced quantum criticality has been reported in many systems, such as in $CoNb_2O_6$ [8], $CeAuSb_2$ [13], $CeCoIn_5$ [14], $YbRh_2Si_2$ [15, 16], $Sr_3Ru_2O_7$ [17, 18, 19], $CePtIn_4$ [31] and many more. Field tuned anomalous quantum criticality has also been observed in Y-substituted $CeNiGe_2$ [20]. Quantum criticality has also been studied in some transition metal oxides. Zn-doping at the V-site in $LiV_2O_4$ results in spin glass as well as metal to insulator transition, documenting that $LiV_2O_4$ is located near to QCP [21, 22]. Moreover, ruthenates of the form $ACu_3Ru_4O_{12}$ (A = Na, Ca, La, Pr, Nd) can be tuned to a QCP, on application of magnetic field or chemical substitution [21]. Hence, it can be said that search for novel systems revealing QCP has gained plenty interest in recent times and is still continuing.

In this context, rare earth orthovanadates, particularly $DyVO_4$, a magnetic insulator, can be quite interesting. However, this compound has not been explored from the viewpoint of quantum criticality. This compound belongs to $RVO_4$ series, in which the indirect exchange between neighboring rare earth magnetic ions may lead to a long-range magnetic ordering at very low temperatures and 4$f$ electron-phonon coupling can result in the lattice instabilities. Members of this series have been extensively investigated in the last few decades as they have been found to exhibit various exotic properties [23 – 27]. Earlier studies on DyVO$_4$ reveal that at ambient conditions, magnetization and $C$ studies gave evidence of the presence of second order long-range AFM ordering at ~ 3 K. However, due to the strong insulating nature of this compound, there are no literature reports about the electrical resistivity. Furthermore, at low temperatures, in the AFM state, DyVO$_4$ undergoes a first order metamagnetic transition (MMT) in an applied magnetic field of ~ 2.1 kOe [26, 27]. Later, neutron diffraction studies confirmed the presence of FOPT from tetragonal to the orthorhombic symmetry at ~ 13.8 K, which was attributed to Jahn-Teller (JT) splitting of a degenerate ground state quartet into two doublets [27]. In the lower symmetry phase, the JT coupling induces a mixing of the two lowest Kramer doublets, leading to large magnetic anisotropy [28, 29, and 30]. This large magnetic anisotropy along with the admixing of eigenfunctions is likely to induce the quantum effects and this increased quantum fluctuations at low temperatures and high magnetic fields might result in the suppression of AFM ordering.

Hence, in this manuscript, we explore the possibility of the existence of the magnetic field tuned QCP in DyVO$_4$ through magnetic, magnetocaloric effect (MCE) and thermodynamic measurements. Our investigations accentuate a perplexing magnetic phase diagram which reveals the presence of quantum TCP separating the paramagnetic (PM), AFM and metamagnetic phase. At 1.8 K, $H$ dependent -d$M$/d$T$, exhibits a minimum followed by a crossover (near 5 kOe) to a maximum. This observation is accompanied by the hyperbolic divergence of d$M$/d$T$ verses $T$ curve at 4.5 and 5 kOe, as $T$ is lowered. These signify the presence of quantum fluctuations along with thermal fluctuations. $C$ curve obtained at 5 kOe shows unusual temperature dependence/slope change; which can be attributed to the deviation from standard behavior of AFM insulators. Further, $C$ and MCE studies indicate towards the presence of entropic topography which shows significant variation as we approach the QCP. Our study suggests that DyVO$_4$ can be tuned to quantum criticality near ~ 2.4 K and ~ 5 kOe.

**Results**

**DC Magnetization**

Fig. 1 (a-b) demonstrates the temperature dependence of the actual magnetic susceptibility ($\chi_a$) at different applied magnetic fields. $\chi_a$ is calculated by considering the concept of demagnetization factor and using the relation ($1/\chi_a = 1/\chi - N$) [43]. Here, $\chi$ and $N$ corresponds to the measured magnetic susceptibility and demagnetization factor, respectively. $\chi_a$ (T) curve at 0.1 kOe is manifested by the presence of a sharp peak at transition temperature ($T_N$) around 3.2 K, which is attributed to long-range AFM ordering. This observation is found to be in good agreement with the report by P.J. Becker *et al* [26]. Further on increasing $H$, the downturn, along with the ordering temperature is decreased, as shown by the arrows in Fig. 1 (a). Above 30 kOe, magnetic saturation with respect to temperature is noted at low temperatures (shown by arrow in Fig. 1 (b)). This observation possibly indicates towards the complete polarization of Dy moments. Further, to understand the field response of $T_N$ (corresponding to $\chi_a$), the variation of $T_N$ with increasing $H$ is plotted in the inset of Fig. 1 (d). The observed decay of $T_N$ is fitted with the power-law of the form

$$T_N (H) = T_N (0) [1 - (H/H_c)^z], \ldots \ldots \quad (1)$$

where $H_c$ is the critical field at which $T_N$ reduces to zero, $T_N$ (0) is the AFM transition temperature at zero field and z is the exponent. The obtained parameters are $H_c = 7.51 \pm 0.33$ kOe, $T_N (0) = 3.21 \pm 0.02$ K and z = $3.34 \pm 0.32$. If $T_N$ (H) smoothly followed the fitted curve, the extracted parameter would suggest that the AFM interaction should cease around 7.5 kOe; implying that a conventional QCP can be achieved around 7.5 kOe. Fig. 1 (c) and 1 (d) focuses on the $\Gamma_{mag}$ and has been discussed later.

Fig. 2 (a) depicts isothermal magnetization curves at different temperature up to 15 kOe in the AFM ordered state (below 3.2 K) as a function of the internal magnetic field ($H_i$). $H_i$ is calculated using the relation: $H_i = (H - NM)$ [30, 43]. One important thing to note here is that before measuring each isothermal magnetization curve, the sample is warmed to a very high temperature (100 K) in the PM region, in order to ensure that the sample is not stuck in previous state before the next measurement. Also, it is noted that $H_i$ shows an increment when $T$ is lowered (shown by an arrow in Fig. S2 of the supplementary information). This is because on lowering the $T$, the magnetic phase below $T_N$ tends to become more stable. At 1.8 K, a sharp change of slope is noted. With an increase in temperature, this change weakens, and vanishes above 2.4 K. This change in slope is believed to be due to the presence of field induced transition in the AFM state, as also reported in literature [27, 30]. It results in the modification in the AFM structure, resulting in a state of strongly staggered moment within the AFM ordered state i.e. the metamagnetic phase. Further, at sufficient high magnetic fields, the $M(H)$ curves saturate and a saturation value $M_s \sim 6.87$ $\mu_B$/Dy is observed at 70 kOe for 1.8 K curve (Fig. 2 (b)). Generally, in polycrystalline samples relatively lower value of the $M_s$, as compared to the theoretically calculated free moment of the magnetic ion, is a signature of presence of strong magnetic anisotropy [41]. In our case, the observed value of $M_s$ is relatively small as compared to the theoretically calculated value ($\sim 10.65$ $\mu_B$/Dy); signifying the presence of strong magnetic anisotropy. This is also in accordance with earlier reports on DyVO$_4$ [27, 30]. Furthermore, this value of $M_s$ is also found to be qualitatively consistent with earlier reports on the isothermal magnetization curves on single crystals as well as polycrystals [26, 30, and 39]. To clearly visualize the change of slope, field response of d$M$/d$H_i$ at selected temperatures in the AFM state is shown in Fig. 2 (c). Here, the maxima corresponding to each curve is ascribed to the MMT [31, 32]. Also, it is noted that the peaks in d$M$/d$H_i$ sharpens and an increase in height is observed as the temperature is decreased. It implies that the MMT becomes stronger at lower temperature. This observation suggests that quantum fluctuations may play an important role in governing the field-driven transitions, as $T \rightarrow 0$ and the tuning parameter $H$ is likely to perturbate these quantum fluctuations [31]. Also, the position of maximum corresponding to the MMT as a function of $H_i$ appears to be sensitive to temperature and therefore, temperature response of critical field associated with MMT demands attention. Hence the maximum value of |d$M$/d$H_i$|$_{max}$ is plotted as a function of temperature (Fig. 2 (d)). The obtained curve is fitted with a power law of the form |d$M$/d$H_i$|$_{max} \sim T^{-n}$, where $n$ is the exponent. As $T$ approaches $T_N$, the experimental points and fitting curve match with each other, while, at low temperatures, a slight deviation is noted (shown by arrow). The obtained value of $n$ is $\sim 2.51 \pm 0.02$. Generally, at zero/low field, the long wavelength fluctuations stabilize a second order AFM transition, and this is reflected in power law divergence of d$M$/d$H_i$ [32]. But the observed unusual deviation at low temperatures in our case can be caused, either by the presence of disorder or due to thermal or quantum fluctuations. The initial assumption can be ruled out as our experimental observations do not provide any evidence of disorder in the system and hence quantum fluctuations (along with thermal fluctuation) are expected to play a role.

We now focus on the thermodynamic quantity, -d$M$/d$T$, as it is an important tool to categorize QCP and to investigate the presence of quantum fluctuations [7, 9]. From the relation,

$$\Gamma_{mag} = -\frac{(dM/dT)_H}{C_H} = \frac{1}{T}\frac{dH}{dT}\bigg|_S \ldots\ldots\ldots \quad (2)$$

it can be inferred that the quantity (-d$M$/d$T$) is related to $\Gamma_{mag}$. For systems having first order line of metamagnetic transition at low temperatures, $\Gamma_{mag}$ develops into a minimum and maximum with a crossover at $H_m$ (or vice versa). The QCP in such systems is described by the end point of the line of first-order transition [9, 10, and 12]. Fig. 1 (c) depicts the $H$ dependent (-d$M$/d$T$) curve at 1.8 K. This parameter first comes close to a negative value forming a minimum at 3.5 kOe and then it approaches ~~to~~ a positive value with a sign change near 5 kOe (~ $H_m$), resulting in a maximum around 6.5 kOe. Generally, for a MMT, a sign change in (-d$M$/d$T$) is expected due to thermal fluctuations. But its evolution to minima and maxima with significant large values hints towards some additional contribution to the fluctuations. Hence, to further support the existence of QPT, the temperature dependent d$M$/d$T$ at 0, 4.5 and 5 kOe is also plotted and shown in the Fig. 1 (d). The curves corresponding to the critical fields show a hyperbolic divergence as the temperature is lowered. Similar kind of behavior had also been reported for $CeRu_2Si_2$ within the metamagnetic state [10, 12]. This phenomenon in $CeRu_2Si_2$ had been ascribed not only to change of antiferromagnetic phase but involves a significant contribution of quantum fluctuations. Hence, it can be said that our observations point towards the possible presence of quantum fluctuations.

**Field dependent AC Susceptibility**

DC field dependent AC susceptibility at different temperatures can be a vital tool to determine the MMT [33, 34] and can also be used to obtain the temperature regime of MMT. In $DyVO_4$, it is expected to show a strong field dependent maximum in the ordered magnetic state, signifying the presence of metamagnetic phase. Field dependent scans of AC susceptibility up to 10 kOe at different temperatures in the AFM state is taken at 331 Hz and at AC field amplitude ($H_{AC}$) of 2 Oe. Fig. 3 (a) shows the real component of AC susceptibility ($\chi'$). At 1.8 K, it exhibits a pronounced peak centered on 4.5 kOe, implying the presence of MMT in the AFM phase. Interestingly, field dependence of the imaginary part of AC susceptibility ($\chi''$) at different temperature, shown in Fig. 3 (b), exhibits some unusual behavior near the vicinity of MMT in the AFM ordered state. It is clearly visible that $\chi''$ curves for $T \leq 2.4$ K start developing as a single maximum centered at $H'' \sim 4.5$ kOe, which is attributed to the dominance of metamagnetic phase at low temperatures. Above 2.4 K, the curves start splitting into two different broad peaks, along with a reduction in the magnitude of $\chi''$. One important point to note here is that this temperature coincides with the temperature above which the slope change feature in the *M* (*H*) disappears, signifying the presence of MMT below 2.4 K. Below 2.4 K, the dynamical response of AC susceptibility in the presence of *H* becomes sensitive to the physics of MMT. Similar behavior has also been reported by W. Wu *et al*., [28]. The double peak feature becomes clearly visible at $T \geq 2.8$ K (i.e., in the AFM state) which can be attributed to the field response of alternating spins in the AFM state. To shed further light on this *H* dependent double peak feature, further investigations are required. Also, from Fig. 3 (b) it is noted that, $H''$ associated with MMT decreases by 0.5 kOe from 1.8 K to 2.4 K. Therefore, from our analysis till now, it can be said that the below 2.4 K, the metamagnetic phase dominates on application of magnetic field, whereas, above this temperature, $DyVO_4$ attains the AFM spin alignment. Further, we have performed MCE which is believed to be an important probe to determine the nature of MMT [13].

**Magnetocaloric Effect**

MCE is the reversible heating and cooling of the magnetic materials upon their exposure to varying magnetic fields [35, 36]. It is an important tool to determine the latent heat associated with the magnetic phase transitions and is also used to study the physics of quantum criticality in the correlated electron systems [37, 38]. A significantly large MCE has already been reported in $DyVO_4$ [39]. In order to investigate the effect of field on AFM and metamagnetic phase, the isothermal magnetic entropy change ($\Delta S$) has been derived from the virgin curves of *M* (*H*) isotherms in the temperature range 1.8-4 K with an interval of 0.2 K using the protocol mentioned in [40]. The sample is warmed every time to PM state before measuring each isotherm. $\Delta S$ is mathematically expressed as

$$\Delta S\ (T, H) = \sum_i \frac{M_i(T+\Delta T) - M_i(T)}{\Delta T} \Delta H_i \qquad \ldots\ldots\ldots \qquad (3)$$

where $M_i$ represents the values of magnetization at temperatures *T* and (*T+ΔT*), for a field change of $\Delta H_i$. Entropy change, - $\Delta S$ between (1.8 - 5) K up to a magnetic field of 10 kOe is shown in the Fig. S2 of the supplementary information and is calculated using the initial condition $T_0 = 1.8$ K and $H_0 = 100$ Oe. This MCE data is converted into a more qualitative information by using $\partial(|\Delta S|/T)/\partial H$ which further can be used to determine the nature of MMT. Generally, for Fermi liquids with a field-independent mass, $\Delta S/T$ is constant. So, it is important to plot its field derivative in order to identify any phase transitions on application of magnetic field [18]. Fig. 4 shows the color scale plot of the calculated $\partial(|\Delta S|/T)/\partial H$ in the *H-T* plane. The low temperature feature (under white curve) around 3-4 kOe below 2.4 K corresponds to large values signifying the entropy shoots (or jumps), confirming the first order nature of MMT within the AFM state. Above this white curve, this value starts reducing, signifying the weakening of MMT, and reinforcing of the spins into the AFM state. At low fields, the region beyond black curve is completely dominated by the AFM state. The transformation from green to blue region at higher fields signifies the complete phase transition from the AFM state to the PM state. The next section focuses on the $T$ and $H$ dependence of $C$ and derived $S$. The entropy is an important tool to determine the QCP as it is expected to exhibit unusual non-linear behavior both for isothermal trajectories varying the tuning parameter $H$ and $T$.

**Heat Capacity**

To further shed light on the intrinsic nature of various field-induced transitions as observed in magnetic and MCE studies, $C$ measurement is performed in different $H$. Temperature response of $C/T$ at 0 Oe up to 40 K is shown in the Fig. 5 (a). The curve shows a sharp peak around 3 K, which is near to the long range AFM ordering temperature. The other maximum observed around 13.4 K, will be discussed in the next section. Fig. 5 (b) shows the magnified $C/T$ vs $T$ curves at different fields up to 6 kOe. One important thing to note here is that the $T_N$ decreases with increasing field and its suppression is terminated around 2.4 K for curves above 3 kOe (shown by red circle), along with the broadening of the transition peak. This termination temperature (i. e. 2.4 K) coincides with the temperature below which the metamagnetic phase dominates, as seen from $\chi''$ ($H$) curves. Alongwith, the magnitude of $C/T$ corresponding to the ordering anomaly decreases and becomes undetectable for higher fields (> 6 kOe) (not shown here). The termination of $T_N$ around 2.4 K for $H \geq 3$ kOe, along with the deviation from power-law hints towards the possible existence of TCP (separating PM, AFM and metamagnetic phase) around 2.4 K, in the critical field regime (3 kOe $\leq H \leq$ 6 kOe).

To map the phase diagram around the expected TCP, magnetic field scans of $C$ is performed at different temperatures in the AFM state (shown in Fig. 5 (c)). Earlier reports on thermodynamics of quantum criticality suggest that $C$ is likely to show a noticeable peak, centered at the position of QCP [7, 18, and 31]. It can be seen from Fig. 5 (c) that the $C/T$ verses $H$ curves exhibit peak-like feature centered at $H_m = 5$ kOe below the TCP (2.4 K). This feature remains consistent as the temperature is lowered, along with the reduction in the magnitude of $C/T$. This appearance of peak in $C/T$ curve is consistent with MMT in the AFM state, as probed by the isothermal magnetization and AC susceptibility. One more remarkable feature of our $C/T$ data is that the position over which the peak is observed remains constant as the temperature decreases. This indicates that this MMT remains consistent in this field regime as the temperature is lowered to 0 K, leading to quantum critical end point (QCEP). Another thing to note here is that, at 1.8 K and below $H_m$, the curves show a power-law like behavior of the form ($\sim (H - H_c)^{-\alpha}$) where, $H_c$ is the critical field corresponding to the position of the peak (i.e., 5 kOe). The red solid line represents the best fit to the power-law behavior. The value is found to be (1.74 ± 0.21). As per literature reports [44], this kind of behavior reflects the divergent behavior of the Gruensein parameter in the neighborhood a field tuned QCP.

Further, the equation (2) suggests that $\Gamma_{mag}$ can be directly accessed by measuring the change in $T$ at constant entropy upon the variation of $H$. It yields the slope of constant entropy curves i.e., isoentropes. An accumulation of entropy is expected near the QCP because of the presence of frustration and inability of the system to choose a suitable ground state [9]. The isoentropes are tilted towards the QCP with a minimum in its vicinity. Again from equation (2), it can be said that $\Gamma_{mag}$ is proportional to the slope of isoentropes and it should have a different sign on each side of the QCP (or in its neighborhood). Fig. 5 (d) shows the isoentropic contour plot in the $H$-$T$ plane below the ascribed TCP. $S$ is calculated by integrating the residual $C_m/T$ (calculated by subtracting the lattice and structural transition contribution) with respect to temperature and is found to be consistent with that calculated from our MCE measurements. The isoentropes have been determined by taking the reference of entropy where it shows a temperature independent behavior at a particular $H$. As seen from the figure, the isoentropes dip to lower and lower temperatures as a function of $H$ which is a signature of quantum criticality. The position of the dip signifies the location of the QCP and is found to be near $H_m$ (= 5 kOe) which is the same field at which a sign change occurs in – (d$M$/d$T$). This gives further validation about the presence of quantum fluctuations. Along with this, the dispersion of the peak centered at QCP into a V-shape (to higher temperature) is a signature of the distortion in the entropy phase diagram [7, 8].

In order to investigate the consequences of the observed quantum fluctuations near 5 kOe, the $T^2$ dependent $C/T$ is plotted at 0, 4.5 and 5 kOe below the TCP (shown in the Fig. 6 (a)). For an AFM insulator, the $C/T$ vs $T^2$ curve shows a linear behavior passing through the origin, implying a negligible electronic contribution to the $C$. Also, the spin-wave contribution varies as $T^3$, and it is indistinguishable from the lattice part [42]. However, in our case, the $T_N$ is quite close to TCP, implying that the contribution from magnons, spin-wave and thermal fluctuations can not be completely neglected and a slight non-linear behavior in the zero field $C/T$ vs $T^2$ curve is expected. From Fig. 6 (a), a small devation from linearity is noted in the curve at 0 kOe. However, for the curves in the critical field regime, a clear change in the curvature with respect to zero field curve is observed which signifies the presence of some additional fluctuations in addition to thermal fluctuations. One thing to note here is that from Fig. 1 (d), showing d$M$/d$T$ as a function of temperature at 4.5 and 5 kOe, also shows divergent behavior as the temperature is lowered, implying the presence of quantum fluctuations. Thus, these additional contributions are believed to be due to quantum fluctuations. Furthermore, to understand the curvature change in the critical field regime, these two curves are fitted with power-law ($y \sim x^{\alpha}$). Here, $y$ and $x$ corresponds to $C/T$ and $T^2$, respectively and $\alpha$ is the exponent and is an important quantity to determine the behavior of $C$ in the critical regime. Generally, for a conventional AFM insulator, the value of this exponent is 1. The obtained value of $\alpha$ for 4.5 kOe and 5 kOe is $0.62 \pm 0.03$ and $0.57 \pm 0.03$, respectvely. It indicates a deviation from standard behavior. Similar kind of argument was also given for hole doped $Eu_2Ir_2O_7$ (an AFM insulator at low temperatures) in which the deviation from standard $C \sim T^3$ behavior at low temperatures is attributed to the presence of quantum fluctuations [42]. Thus, the unsual behavior in $C$ is ascribed to the presence of quantum critical fluctutaions near 5 kOe. Hence our investigation suggests to the existence of field induced quantum criticality in $DyVO_4$, along with QCP around 5 kOe, thereby, giving the information about the distortion of phase diagram.

**Discussion**

As mentioned in the section 2, $DyVO_4$ belongs to Zircon class of minerals, crystallizing in the tetragonal crystal structure. Also, from Fig. 5 (a) it is noted that the curve shows a broad maximum at $T_D = 13.4$ K, followed by a sharp peak around 3 K. Earlier neutron diffraction study [27] suggests that the feature at higher temperature can be attributed to a first order JT structural transition from tetragonal to orthorhombic symmetry. Optical spectroscopy investigation on $DyVO_4$ [28, 30] also suggests that in tetragonal symmetry, the $Dy^{3+}$ ($4f^9$) ions with ground state term $6_{H_{15/2}}$ are Kramers ion. The crystal field of the tetragonal symmetry splits the $6_{H_{15/2}}$ manifold into eight Kramer doublets. The two lowest doublets are separated by $\Delta = 13.7$ K and the pairs are almost degenerate. Below $T_D$, the orthorhombic crystal field driven by JT coupling results in the admixing of the eigenfunctions of two lowest

Kramers doublets. As a consequence of this, the separation between them is increased to $\Delta' = 39$ K. This argument is further accompanied by large magnetic anisotropy as seen in the calculated g-factors [29, 30] which is supported by our analysis of relatively low value of $M_s$ discussed in section 3.1. The above two factors are expected to induce quantum fluctuations which are most likely to dominate at low temperatures and with increasing fields. Results obtained from our experimental studies suggest a sharp rise in magnetization. It occurs due the breaking of AFM spin alignment on applying field; resulting, in a state of higher magnetization. Further, this state is found to coincide with the AFM and PM phases at TCP. In order to precisely determine the boundaries of these magnetic phases and track their dependence on the external stimuli (i. e. *T* and *H*), we have constructed a magnetic phase diagram displayed as *C*/*T* contour plot in *H-T* plane (Fig. 6 (b). In the diagram the boundary of the short-range PM region is extracted from *C*/*T* verses *T* data at 0 kOe. In a tetragonal symmetry, each $Dy^{3+}$ ion is surrounded by four nearest $Dy^{3+}$ neighbors. At 13.4 K, the structural transition reduces the four nearest neighbors (NN) to a pair of two NNs, which may probably affect the magnetic interactions. This limited number of NNs in orthorhombic phase is believed to give rise to the short-range PM type of correlations in the phase I. Similar kind of argument is also reported from the optical investigation of $DyVO_4$ [27, 29]. The low field region displaying high values of *C*/*T* is the AFM region of this compound. On application of *H*, $T_N$ decreases signifying the suppression of AFM region with increasing *H* (shown by open circles). The red dash curve in Fig. 6 shows the power-law fitting behavior separating the AFM and PM phases. For this compound, power analysis suggested that the conventional QCP should have been achieved around 7.5 kOe (also shown in Fig. 1 (c). However, the conventional criteria cannot be followed in $DyVO_4$ as the observed second order phase line terminates at the TCP. Since magnetic ordering is associated with a broken symmetry, the phase line must continue to $T_N = 0$ K as a line of FOPT, which is then terminated at quantum critical end point (QCEP). This argument matches well with our results. In $DyVO_4$, field response of *C*/*T* shows a suppression of the second order AFM transition which terminates at TCP (~ 2.4 K) shown by the black arrow. Below TCP, the spins transform into a state of strongly staggered moments, described as metamagnetic state. The first-order nature of MMT is confirmed by the entropy change color plot deduced from MCE data which agrees with earlier experiments [30]. As seen from the phase diagram, below TCP (2.4 K), the *C*/*T* contour lines show sharp dip like behavior as *T* is lowered, centered at 5 kOe. This implies that at 2.4 K, the second order AFM phase develops into first order MMT which continues in *H-T* plane as *T* decreases, leading to the QCEP. Further, the power-law dependence of *C*/*T* ($H$) at low fields for 1.8 K curve reflects the divergence of the $\Gamma_{mag}$ in the vicinity of field induced QCP. The isoentropic curves show dip like behavior centered at $H_m$ and the curves dip to lower and lower $T$ as a response to $H$ which is a signature of quantum criticality. The mapping out of the peak, seen at QCP in the $S/T$ entropic topography in the form a V-shaded quantum critical region signifies a distortion of the phase diagram. Furthermore, the evolution of minimum and maximum of the quantity (-d$M$/d$T$) with a crossover near 5 kOe along with the hyperbolic divergence observed in $T$ dependent d$M$/d$T$ at the same field as $T$ decreases, give hint of significant contribution of quantum fluctuations. This argument is further supported by the unusual $T$ dependence observed in $C$ around 5 kOe. Hence, all our experimental observations and analysis suggest the possibility of formation of QCP at 5 kOe with TCP around 2.4 K. Thus, it can be inferred that in a magnetic insulator DyVO$_4$, quantum criticality is achieved using $H$ as a tuning parameter, with a distorted entropy $H$-$T$ phase diagram at QCP.

**Conclusion**

In conclusion, we have systematically tuned the magnetic and thermodynamic properties of DyVO$_4$, to explore the possibility of quantum criticality through entropic topography. Application of magnetic field results in the suppression of second order AFM transition. This transition is terminated at TCP, where PM, AFM and metamagnetic phases coincide. Beyond this point, the phase line continues as a first order MMT, as $T$ approaches zero. The quantity d$M$/d$T$ and $C/T$ vs $T^2$ curves shows the evidence of quantum critical fluctuations and deviation from conventional behavior near the QCP. Entropic topography in $H$-$T$ plane gives hint about the distortion of the phase diagram creating a V-shaped quantum critical region. Results on DyVO$_4$ indicates that it is a potential candidate to study the mystifying physics behind the distortion of phase diagram, at QCP in insulating magnetic materials. Our investigation might initiate further studies on similar kind of systems in the context of quantum criticality, by using additional tuning parameter like external pressure or angle dependence of the applied field.

**Methods**

Polycrystalline sample of DyVO$_4$ has been prepared by the conventional solid state route by using the precursors Dy$_2$O$_3$ and VO$_2$ from Sigma Aldrich, with 99.99 % purity. The initial reagents were taken in the stochiometric ratio, grinded and given first heat treatment at 800º C, followed by final sintering at 900º C. Room temperature powder X-ray diffraction (XRD) was performed at room temperature in the range 10°-90° using Rigaku diffraction with Cu Kα (λ = 1.54). FullProf Suite software was used to perform the refinement of the X-ray diffraction data. Fig. S1 of the supplementary information shows the Rietveld refinement of the XRD pattern. The compound belongs to Zircon family, crystallizing in the tetragonal structure with space group *I 41/a m d* and is in single phase. *T* and *H* dependent magnetic measurements were performed using the Magnetic Property Measurement System (MPMS) from Quantum Design, USA. Physical Property Measurement System (PPMS), Quantum Design, USA was used to measure field dependent *C*.

**Acknowledgements**

The authors acknowledge IIT Mandi for the experimental facilities and financial support.

# Figures

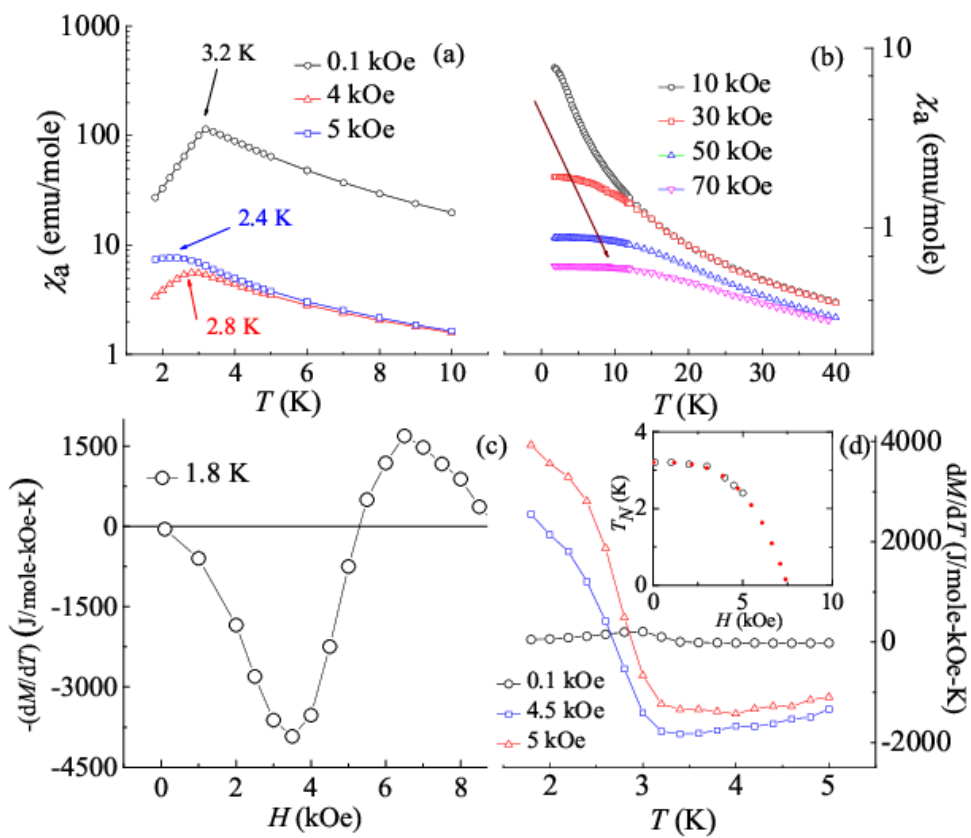

**Fig. 1** (a-b) $\chi_a$ as a function of temperature at different applied magnetic fields. In (a) the arrows represent transition temperature ($T_N$) at various fields whereas in (b) the arrow shows the temperature corresponding to the saturation of $\chi_a$ ($T$). (c) $H$ response of the thermodynamic quantity $-(dM/dT)$ at 1.8 K, showing a sign change just above 5 kOe. (d) $T$ dependence of $dM/dT$ at various applied external magnetic fields. The inset shows the behavior of $T_N$ as a response to the applied field. Red dashed line corresponds to the power law fitting.

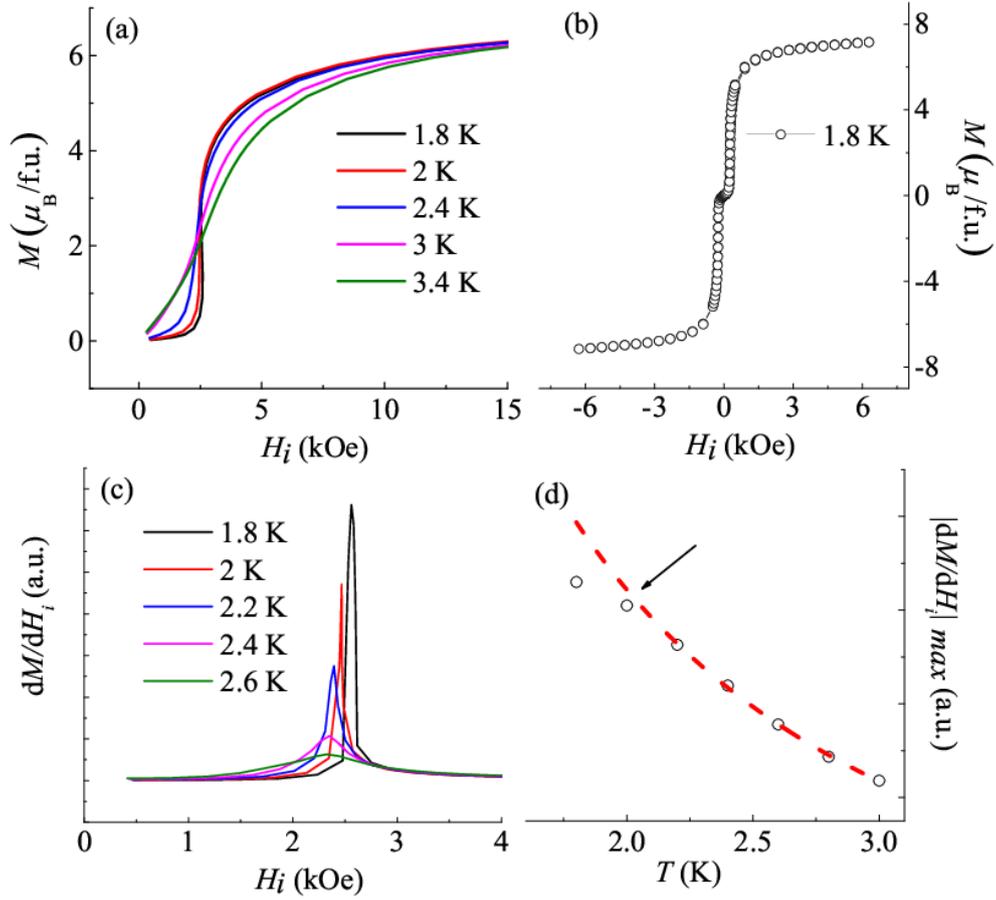

**Fig. 2.** (a) $M(H_i)$ curves up to 15 kOe at various temperatures in the AFM ordered state. (b) $M(H_i)$ curve at 1.8 K up to 70 kOe. (c) $dM/dH_i$ at different temperatures in the AFM ordered state plotted as a function of internal field. (d) Temperature dependence of the maximum value in differential susceptibility, $|dM/dH_i|_{max}$, plotted as a function of temperature. Red dashed line represents the power-law fitting of the form $T^{-n}$. The black arrow denotes the deviation from power-law at low temperatures.

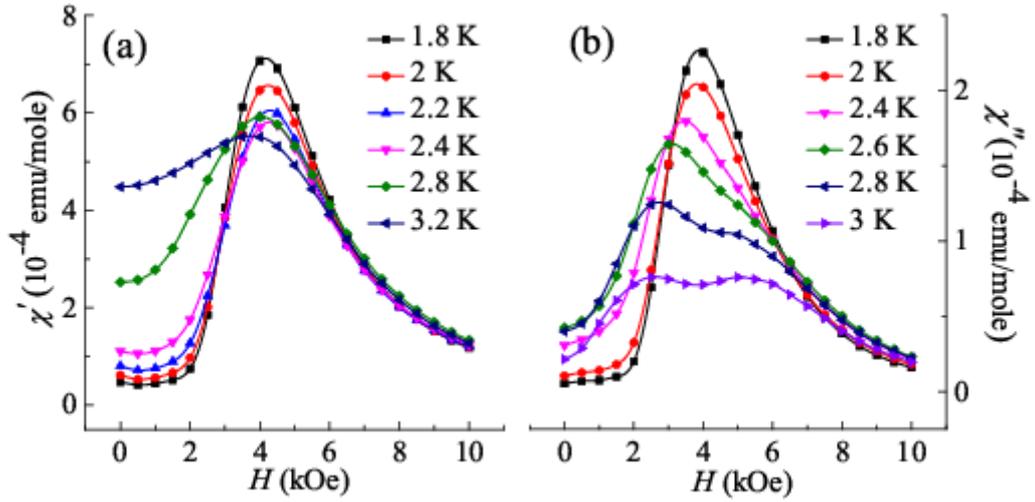

**Fig. 3** (a, b) Real ($\chi'$) and imaginary ($\chi''$) part of the AC susceptibility (AC field = 2 Oe, and frequency = 331 Hz) at various temperatures below $T_N$, plotted as a function of applied DC field.

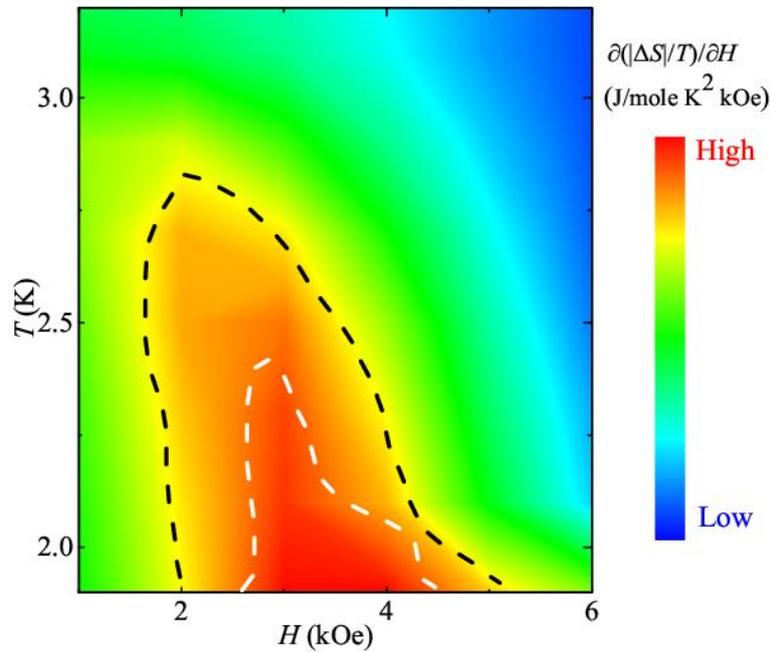

**Fig. 4.** Entropy change ($|\Delta S|$) landscape color map plot derived from MCE data. The white and black dashed curves in the figure separate different regions.

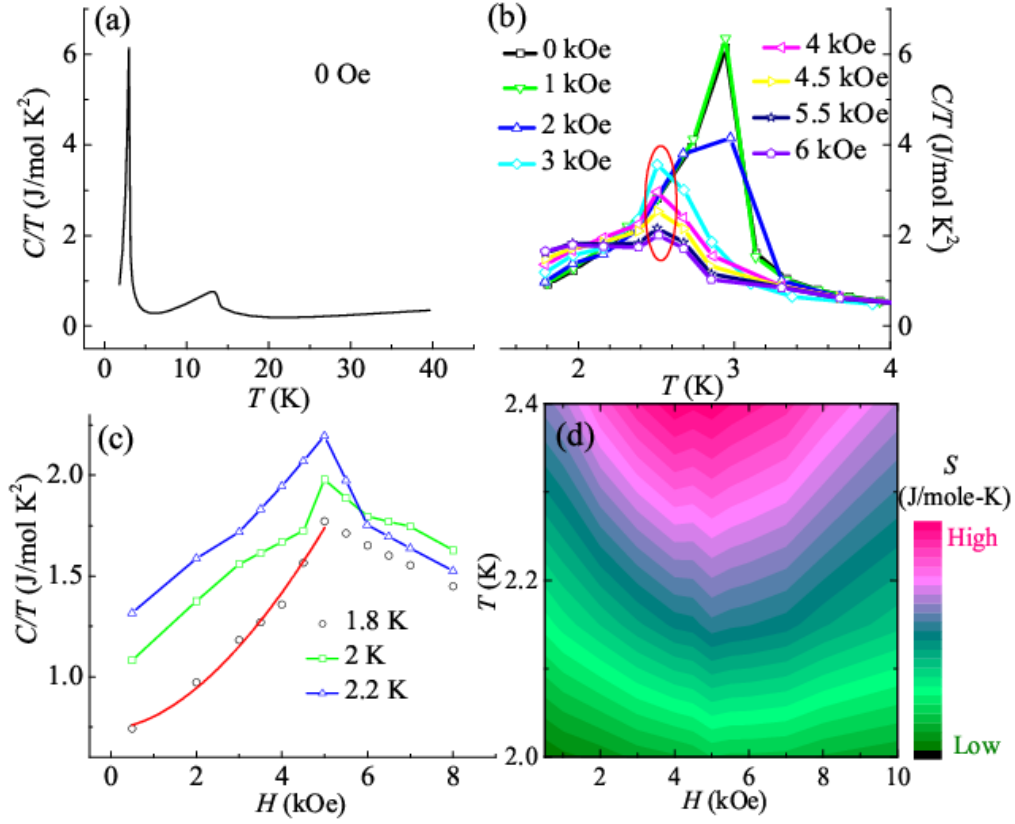

**Fig. 5.** (a) *C/T* at 0 Oe plotted as a function of temperature up to 40 K. (b) Similar plot at different applied field up to 4 K. The red circle in the figure represents the region where suppression of AFM ordering is terminated. (c) *C/T* at different temperatures below TCP, plotted as a function of applied field. The red solid line shows the fit to power-law behavior ($\sim (H - H_c)^{-\alpha}$). (d) Isoentropes dip to lower and lower temperature as a function of applied magnetic field.

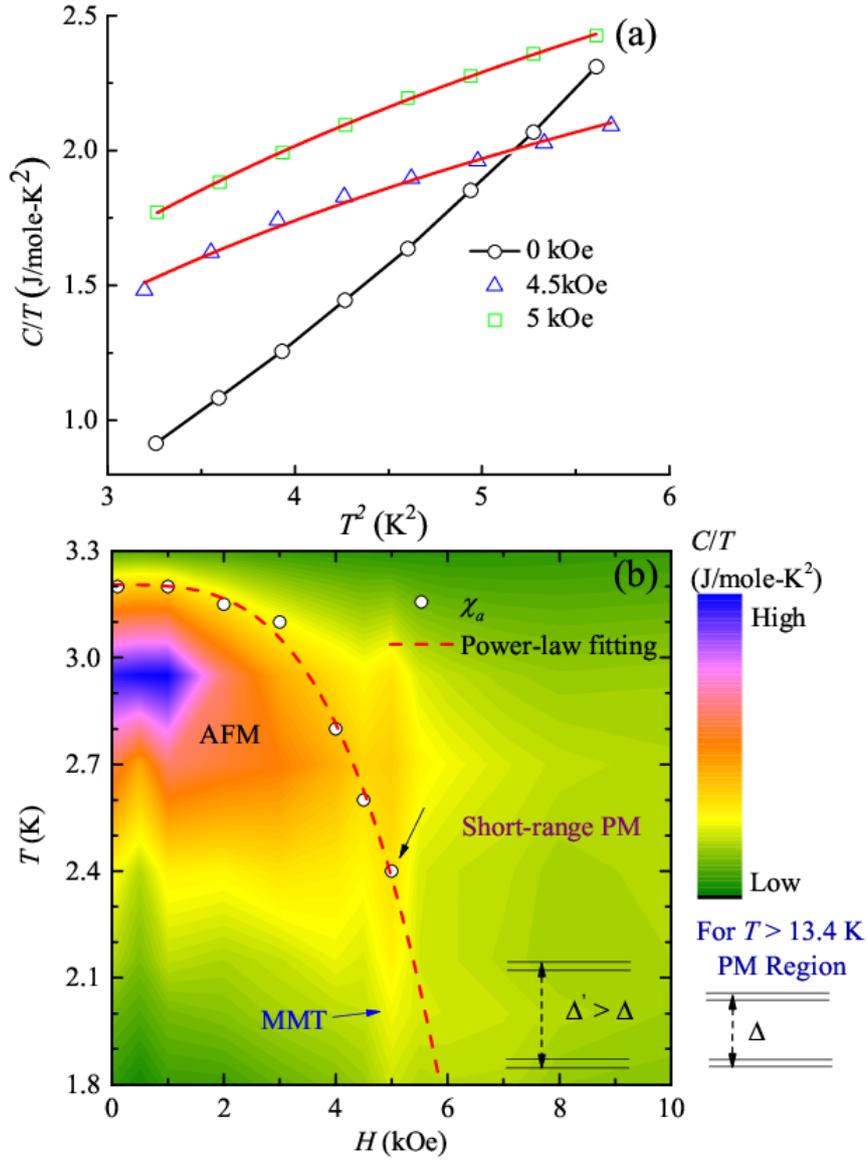

**Fig. 6** (a) $C/T$ vs $T^2$ curves obtained at 0 kOe, 4.5 kOe and 5 kOe. Solid red lines show the power-law fitting to the data. (b) Magnetic phase diagram for DyVO$_4$ presented as an $H$-$T$ color map plot with $C/T$ represented as the color scale. The red solid line corresponds to the power-law dependence of $T_N$. The black arrow shows the position of TCP.

# Supplementary Information

# Entropic topography associated with field-induced quantum criticality in an antiferromagnet $DyVO_4$

Dheeraj Ranaut and K. Mukherjee

School of Basic Sciences, Indian Institute of Technology Mandi, Mandi 175005, Himachal Pradesh, India

1. **Crystal Structure**

   Rietveld refinement of the XRD pattern obtained at 300 K shows that $DyVO_4$ crystallizes in the tetragonal structure with space group *I 41/a m d* and is in single phase (Fig. S1).

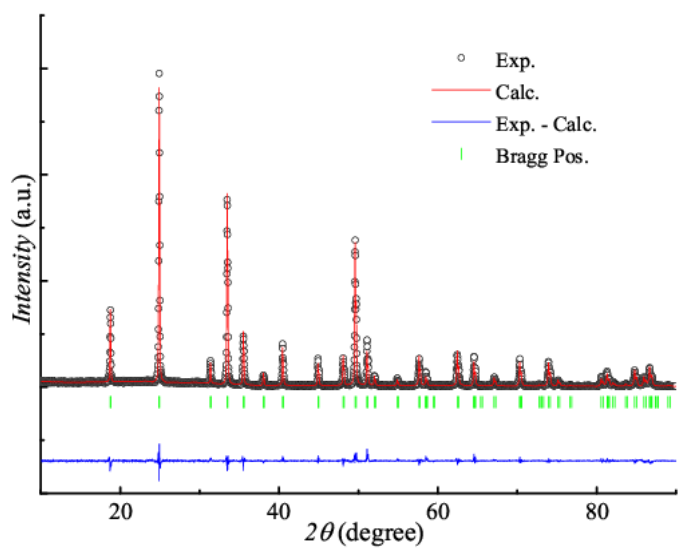

**Fig. S1** Rietveld refined powder x-ray diffraction pattern of $DyVO_4$ at 300 K. The open circles indicate the experimental data, while the Ritveld refined pattern is shown by red solid line. The difference curve and Bragg positions are denoted by blue line and green vertical bars, respectively.

2. **Internal magnetic field ($H_i$) measurements:**

   In case of samples with large magnetization, the demagnetization factor needs to be taken in account in order to gain more insight into the field induced magnetic transitions. In polycrystals, generally field induced transitions appear very broad and not much intrinsic information can be gathered. Taking the demagnetization factor results in very sharp transitions, giving more intrinsic information about the system [Ref. 30]. Therefore, in order to get more intrinsic information of the effect of applied magnetic field on magnetic transitions, we have considered demagnetization factor. Thus, the actual magnetic susceptibility ($\chi_a$) and internal magnetic field ($H_i$) has been calculated using the value of demagnetization factor with following relations:

$$H_i = H - NM \ldots\ldots (1)$$

where, $H_i$ is the internal magnetic field experienced by the sample, $H$ is the applied external magnetic field, $N$ and $M$ are the demagnetization factor and magnetization, respectively.

Hence, we can write:

$$1/\chi_a = 1/\chi - N \ldots\ldots (2)$$

where, $\chi_a$ (=$M/H_i$) is the actual magnetic susceptibility and $\chi$ (=$M/H$) is the measured magnetic susceptibility.

Fig. 1(a), 1 (b), 2 of the main manuscript are plotted with respect to $\chi_a$ and $H_i$. Further, it is noticed that within the AFM state, $H_i$ increase as $T$ is lowered; as at lower temperatures the magnetic phase becomes more stable. Fig. S2 illustrates different $M$ ($H_i$) curves at different $T$ within the AFM state and it is clearly visible that $H_i$ increases as $T$ decreases (shown blue the blue solid arrow).

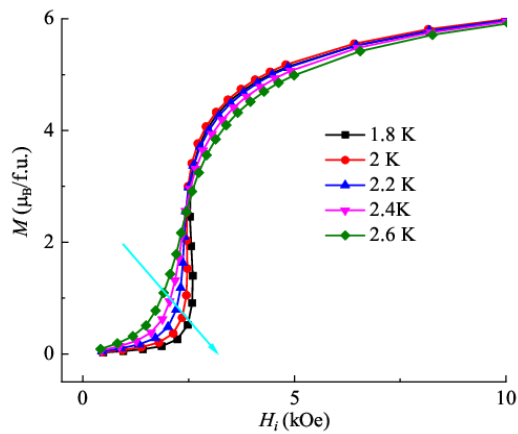

**Fig. S2.** $M$ ($H_i$) curves at different temperatures within the AFM state. The blue arrow shows that as temperature is lowered, the $H_i$ increases.

## 3) Magnetocaloric Effect (MCE)

MCE in temperature range of 1.8 - 5 K and up to a field of 10 kOe, is shown in Fig. S2. An interesting observation is the effect of field on the -$\Delta S$, below the $T_N$. It is noted that below 3 K, the magnitude of the MCE first increases with field up to 5 kOe and then starts decreasing with increasing field. This unusual behavior within the AFM state is believed to be due to the presence of metamagnetic transition with a critical field around 5 kOe. Also, it is observed that the curves become more prominent at higher field around the AFM transition.

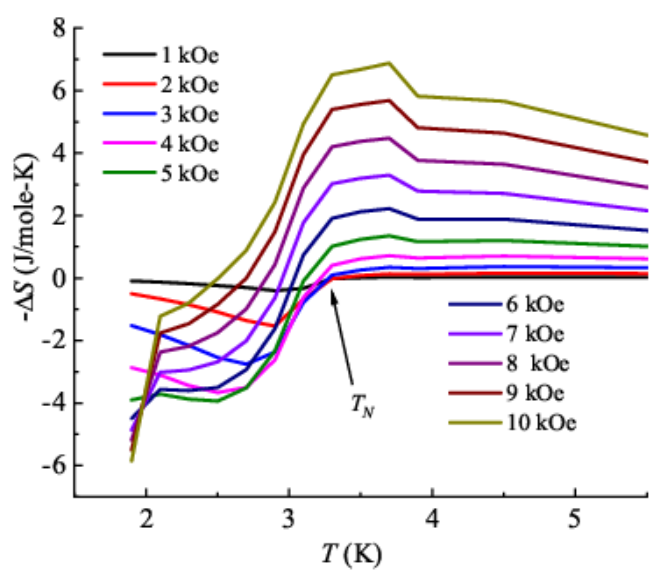

**Fig. S2.** Entropy change, **-**$\Delta S$ plotted as function of temperature (below 5 K) for magnetic fields (1–10 kOe). The arrow shows the transition temperature.